\documentclass[sigconf]{acmart}

\usepackage{balance}

\usepackage{subfig}
\usepackage{booktabs} 
\usepackage[english]{babel}
\usepackage{array}
\usepackage{tabularx}
\usepackage{relsize}
\usepackage{rotating}
\usepackage[export]{adjustbox}
\usepackage{multirow}
\usepackage{listings}
\usepackage{amsmath}
\usepackage[ruled,vlined,linesnumbered]{algorithm2e}
\usepackage{graphicx}
\graphicspath{ {./images_no_type3_fonts/} }

\def\BibTeX{{\rm B\kern-.05em{\sc i\kern-.025em b}\kern-.08emT\kern-.1667em\lower.7ex\hbox{E}\kern-.125emX}}

\copyrightyear{2019}
\acmYear{2019}
\setcopyright{acmlicensed}
\acmConference[KDD '19]{The 25th ACM SIGKDD Conference on Knowledge Discovery and Data Mining}{August 4--8, 2019}{Anchorage, AK, USA}
\acmBooktitle{The 25th ACM SIGKDD Conference on Knowledge Discovery and Data Mining (KDD '19), August 4--8, 2019, Anchorage, AK, USA}
\acmPrice{15.00}
\acmDOI{10.1145/3292500.3330689}
\acmISBN{978-1-4503-6201-6/19/08}

\acmSubmissionID{ads0576p}

\begin{document}
\fancyhead{}
\title{Real-time Event Detection on Social Data Streams}
\author{Mateusz Fedoryszak}
\affiliation{
  \institution{Twitter}
  \city{London}
  \country{UK}
}
\email{mfedoryszak@twitter.com}
\authornote{Equal contribution}

\author{Brent Frederick}
\affiliation{
  \institution{Twitter}
  \city{New York}
  \country{USA}
}
\email{brentf@twitter.com}
\authornotemark[1]

\author{Vijay Rajaram}
\affiliation{
  \institution{Twitter}
  \city{New York}
  \country{USA}
}
\email{vrajaram@twitter.com}
\authornotemark[1]

\author{Changtao Zhong}
\affiliation{
  \institution{Twitter}
  \city{London}
  \country{UK}
}
\email{czhong@twitter.com}
\authornotemark[1]

%

\begin{abstract}
Social networks are quickly becoming the primary medium for discussing what is happening around real-world events. The information that is generated on social platforms like Twitter can produce rich data streams for immediate insights into ongoing matters and the conversations around them. To tackle the problem of event detection, we model events as a list of clusters of trending entities over time. We describe a real-time system for discovering events that is modular in design and novel in scale and speed: it applies clustering on a large stream with millions of entities per minute and produces a dynamically updated set of events. In order to assess clustering methodologies, we build an evaluation dataset derived from a snapshot of the full Twitter Firehose and propose novel metrics for measuring clustering quality. Through experiments and system profiling, we highlight key results from the offline and online pipelines. Finally, we visualize a high profile event on Twitter to show the importance of modeling the evolution of events, especially those detected from social data streams.
\end{abstract}

%
%


\maketitle

\section{Introduction}
Social networks are being increasingly used for news by both journalists and consumers alike. For journalists, they are a key way to distribute news and engage with audiences: a 2017 survey \cite{cision} found that more than half of journalists stated social media was their preferred mode of communication with the public. In addition, journalists also use social media frequently for sourcing news stories because they ``promise faster access to elites, to the voice of the people, and to regions of the world that are otherwise difficult to access'' \cite{von2018sourcing}. For consumers, according to a recent Pew survey, social networks have surpassed print media as their primary source of news gathering \cite{adweek}. Factoring in journalists' predilection for breaking stories on social media, audiences often turn to social networks for discovering what is happening in the world.

Thus, understanding the real-time conversation from both journalists and audiences can give us rich insight into events, provided we can detect and characterize them. Before diving into how we analyze conversation to identify events, let us first define an event. Previous work by McMinn et al. \cite{mcminn2013building} gave a broad definition of an event as a ``significant thing that happens at some specific time and place.'' What the authors argued was that something ``significant'' is happening when ``it may be discussed in the media.'' Furthermore, they stated that events are representable by the group of entities people use to talk about the event. For example, an event for a film awards show can be represented by the nominated actors, actresses, and films that are being discussed. 

We propose extending this definition in two ways. First, we argue a significant thing is happening when a group of people are talking about it in a magnitude that is different from normal levels of conversation about the matter, or in other words, it is trending. Second, we claim that this eventful conversation can change over time, and our data model for an event should reflect this. Thus, we model an event as a list of clusters of trending entities indexed in time order, also referred to as a cluster chain.  A detected event corresponds to a cluster chain and is characterized at a particular point in time by a cluster of trending entities.

We apply this definition to the problem of event detection from Twitter data streams. Twitter has unique characteristics that we believe make the problem particularly challenging. The first of these is the scale. There are approximately 500 million tweets a day (or $6K$ tweets per second on average on the complete Twitter Firehose - the stream of all tweets). The second is the brevity. Semantic understanding of text is difficult given that tweets are written in a unique conversational style particular to the brevity of the Twitter medium (280 character limit per tweet). The third is the noise. Many of the tweets on the platform are unrelated to events and even those that are related can include irrelevant terms. The fourth and final characteristic is the dynamic nature of what is discussed on the platform. Event detection can not be static: we have to track the evolution of events over time and handle continuity and discontinuity in conversation. 

To address these challenges, we built a real-time system to ingest the full Firehose and identify clusters of event-related entities on a minute-by-minute basis. We link these clusters into cluster chains as the event progresses over time. We attempt to quantify and reduce the impact of noise by creating an offline simulation framework that allows us to experiment with methodologies in order to produce high-quality events. As part of this offline pipeline, we extract a set of candidate events from a day's worth of Twitter data, along with their associated clusters of entities in an evaluation dataset. Relying on the optimal clustering methodology derived offline, we implement it in an online system that can account for the scale and fluctuations of the Twitter stream. Our observation is this dual mode approach of online and offline systems is complementary and allows for the separation of concerns: the former is optimized for low latency and scalability while the latter is focused on assessing various methods and attaining high quality.

This work has resulted in a production application that solves various product use cases and improves metrics related to user exploration of ongoing events. 
Compared to previously published work on event detection, this paper has several novel contributions:
\begin{enumerate}
\item \emph{Tracking of event evolution over time} - Based on our review of the literature, relatively little attention has been given previously to this subject. We argue that representing an event as a chain of clusters over time is a powerful abstraction. Moreover, we are able to track these cluster chains in real time. Temporal analysis of clusters yields insights about sub-events and audience interest shifts. We highlight a case study of a high profile event on Twitter to demonstrate this. 
\item \emph{Differentiated focus on quality of clustering} - We introduce new metrics for entity clustering quality that we believe can help ground subsequent efforts in the space. Through quantitative experiments, we demonstrate the trade-offs of key system parameters and how they impact quality and coverage. 
\item \emph{Novel real-time system design} - Previous work that operated on large-scale Twitter data in real time \cite{mcminn2015real} combined detection and clustering of bursty terms in a sequential pipeline. Our design is based on the realization that we can decompose burst detection and clustering into separate components that can be scaled independently. Through system profiling, we demonstrate the scalability and resilience of this approach. 
\end{enumerate}

\section{Related Work}
Depending on the context, there are varying definitions of the problem of event detection. In the context of newswire documents, Orr et al. \cite{orr2018event} framed event detection as identifying ``trigger words'' and categorizing events into ``refined types.'' However, with respect to social data streams, it is difficult to predict trigger words, given the unstructured and noisy nature of the documents.

Despite these challenges, due to its public nature, Twitter is used as the source of data in various event discovery research projects focusing on social data. In the context of Twitter, McMinn et Jose \cite{mcminn2015real} approached event detection as clustering a stream of tweets into the appropriate event-based cluster. Guille and Favre \cite{guille2015event} instead clustered relevant words from the stream of tweets. The choice of whether to cluster terms or documents in order to identify events is evident in the literature.

The most significant event detection techniques have been surveyed by \cite{atefeh2015survey, hasan2017survey}. These techniques can be broadly categorized as either feature-pivot (FP) \cite{fung2005parameter} or document-pivot (DP) \cite{yang1998study} methods. The former corresponds to grouping entities within documents according to their distributions while the latter entails clustering on documents based on their semantic distance \cite{li2017real}.

Fung et al. \cite{fung2005parameter} argued feature-pivot methods are easier to configure because they contain fewer parameters than document-pivot methods. Another difference is DP-based clustering results in potential additional work to summarize the events: one has to generate a summary from the tweets (e.g. by selecting the top tweet) whereas with FP, the list of entities is a condensed representation that can serve as a summary. Conversely, in order to find the best tweets for an event, a search query has to be generated from entities extracted in FP whereas with DP, the tweets are already present. Both methods are widely employed for event detection with Twitter data and have been proven to be effective. To minimize parameter tuning and serve various product use cases, entity clustering rather than tweet clustering was selected in this work. Consequently, we employ a FP technique.

One popular class of FP techniques is topic detection, which attempts to identify events by modeling documents as ``mixtures of topics, where a topic is a probability distribution over words'' \cite{stilo2016efficient}. As the event changes over time and people use different words to discuss it, the probability distributions of the underlying topic representations change. However, as \cite{stilo2016efficient} points out, it is difficult to capture ``good'' topics from short documents such as tweets; moreover, these approaches are susceptible to memory problems with high volume datasets or topic counts, which are often produced in a large-scale production environment. Finally, many topic detection approaches do not adequately capture the ``burstiness,'' or velocity, of words over time, and this is critical to distinguish events from non-events \cite{stilo2016efficient, li2014online}.

Bursty terms on Twitter are defined as those appearing in an unusually high rate of tweets. Numerous studies have attempted to leverage bursty term tracking for event discovery. For example, TwitterMonitor \cite{mathioudakis2010twittermonitor} performed event detection by identifying bursty words and then merging them into groups based on their co-occurrence in tweets using a greedy algorithm. Each group represents an event. This is similar to our approach; however, we do not rely on a greedy selection of co-occurrences. We instead track all co-occurrences over a time window. EDCoW \cite{weng2011event} followed this process but used wavelet decomposition to identify bursty words.

Most of the methods above fail to consider the evolution of events. The importance of temporal evolution is discussed in event visualization design \cite{dork2010visual, guille2015event}. For example, Archambault et al. \cite{archambault2011themecrowds} highlighted an example of the tsunami in Japan that occurred in March 2011. Initially, the event is dominated by keywords like ``earthquake'' and ``tsunami'' but later words such as ``nuclear'' and ``radiation'' are introduced.

Some recent studies \cite{petrovic2010streaming, osborne2014real, becker2011beyond} proposed using incremental clustering \cite{hasan2016twitternews} to solve the event evolution problem. Models are incrementally updated as new data arrives on a stream. Such methods may not be feasible to use for the Twitter Firehose due to the scale of updates. In this paper, we solve this problem by adding a layer of cluster linking into a FP method using an idea similar to evolutionary clustering \cite{chakrabarti2006evolutionary}. This type of linking was proposed in \cite{lee2014incremental}; however, they were not able to demonstrate their end-to-end approach performing in an online setting. Our approach achieves event detection with evolution tracking in real time through modeling events as cluster chains and addressing scaling concerns with new design choices.

\section{Methodology}

\begin{figure*}[h]
\includegraphics[width=0.9\textwidth, keepaspectratio]{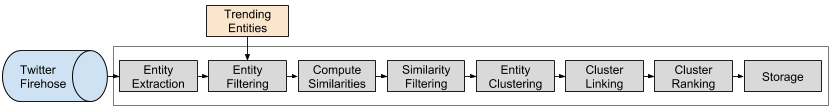}
\caption{Clustering service design}
\label{fig:ClusteringService}
\end{figure*}

\subsection{Framework}

Figure \ref{fig:ClusteringService} summarizes the end-to-end framework to output entity clusters from a stream of data. One key advantage of the framework is the modular composition. As a result, we can test components in isolation and replace algorithms at appropriate parts of the pipeline in order to improve the overall output. To describe this system, we first start with the important terminology and then delineate each of the components.

\subsubsection{Terminology}
\begin{itemize}
	\item Entity - A tag for some content (e.g. text, image) in a document (e.g. tweet). Examples used in this work include named entities \cite{namazifar2017named} and hashtags but we can extend to other entity types such as user IDs or URLs.
	\item Cluster - A set of entities and their associated metadata (e.g. entity frequency count)
	\item Cluster Chain - A list of clusters over time that is related to the same ongoing event
	\item Event - A cluster chain along with any metadata (e.g. detected start time, end time)
\end{itemize}

\subsubsection{Trend Detection}
\label{section:TrendDetection}
In our method, we focus on clustering entities that are trending \cite{trends, instagram} by leveraging an internal trend detection system known as Twitter Trends. By ingesting the input of the Firehose, the Twitter Trends service computes trending entities across geographical locations. It does so in real time via a Summingbird topology \cite{summingbird} and has the following key phases:
\begin{itemize}
	\item Data preparation - This step includes filtering and throttling. Basic filtering removes Tweets with low text quality or sensitive content. ``Throttling removes similar Tweets and ensures contribution to a trend from a single user is limited.'' \cite{trends}
	\item Entity, domain extraction and counting - For a given tweet, we extract the available entities and geographical domains. For every domain and entity, we emit a count with a tuple of $<entity, domain, 1>$ and aggregate this over time.
	\item Scoring - The scoring is based on anomaly detection: we compute expected $<entity, domain>$ counts and compare that with observed counts to generate a score for each pair. To calculate expected count for a domain and entity pair, we use the following formula:
	\begin{equation}
	E(d,e) = \frac{N_{s}(d)}{N_{l}(d)} \cdot N_{l}(d,e)
	\end{equation}
	where $E(d, e)$ is expected count for domain $d$ and entity $e$, $N_{l}$ is count over a long time window and $N_{s}$ is count over a short window.
	\item Ranking - The top scoring trends per domain are persisted and available to be queried.
\end{itemize}

The data preparation stage serves to combat the noise and redundant information from the stream. By identifying trending entities, we are able to derive signal from each tweet despite their brevity. More details are provided in \cite{trends}. As mentioned earlier, prior approaches have combined trend detection with clustering by sequentially composing the steps. We propose instead to make the components asynchronous from one another. This separation aligns well with our microservices based architecture: we have a trend detection service and a clustering service and it allows us to scale each independently. Decoupling them also enables us to iterate and enhance their capabilities separately.

\subsubsection{Entity Extraction}
Here are some example entity types that are extracted from each tweet:
\begin{itemize}
\item Named entities - e.g. ``Oprah Winfrey''
\item Hashtags - e.g. ``\#UsOpen''
\item Internal knowledge graph entities - e.g. ``ENTITY 123''
\end{itemize}
At this stage, we process entities all tweets that have made it past any initial filters. Note that our implementation can easily extend to multiple entity types.
\subsubsection{Entity Filtering}
Filtering is extensible but we mainly employ trending entity filtering. By periodically querying the trends service (refer to \S \ref{section:TrendDetection}), we cache the latest set of trends and use it to filter out non-trending entities.
\subsubsection{Compute Similarities}
With the remaining filtered entities, we track their frequency count and co-occurrences amongst them over a sliding time window $W$. We use these frequencies and co-occurrences to compute similarities between entities. Let us take the following example tweets to illustrate further:

\begin{table}[h]
\centering
\begin{tabular}{c|l}
  TweetID & Text  \\
  \hline
  1 & iphone released during \#appleevent\\ \hline
  2 & Tim Cook presents the new iphone \#appleevent\\ \hline
  3 & Tim Cook unveiled the iphone\\
  \hline
\end{tabular}
\caption{Example tweets}
\vspace{-0.8cm}
\label{table:Tweets}
\end{table}
We can represent the co-occurrences for entities as seen below:
\begin{table}[h]
\centering
\begin{tabular}{@{\extracolsep{4pt}}l|c|c|c}
& tweet1 & tweet2 & tweet3  \\
\hline
iphone & \textbf{1} & \textbf{1} & \textbf{1} \\ \hline
\#appleevent & \textbf{1} & \textbf{1} & 0 \\
\hline
\end{tabular}
\caption{Encoding of entities}
\label{table:Encoding}
\vspace{-0.8cm}
\end{table}

The entity vectors for $iphone$ and $\#appleevent$ are the corresponding rows:
\begin{align*}
iphone = [1, 1, 1] \\
\#appleevent = [1, 1, 0]
\end{align*}
Cosine similarity for two entities X and Y:
\begin{equation}
\cos(X, Y) = \frac{X \cdot Y}{\| X \| \| Y \|}
\end{equation}
For example,
\begin{align*}
cos(iphone, \#appleevent) = 0.81649
\end{align*}

The potential disadvantage of this type of encoding is that it gets extremely sparse as we process more tweets; we avoid this by densifying the representation needed to update entity co-occurrences and frequencies. We observe that this type of cosine similarity works well in practice with respect to the final clustering output (see Evaluation section).
\subsubsection{Similarity Filtering}
Once we compute entity similarities, we can filter them based on the minimum threshold $S$ to remove noisy connections between entities.
\subsubsection{Entity Clustering}
At this stage, we are able to naturally construct a graph consisting of the entities as nodes and their similarities as edge weights. Once we can compute similarities, the advantage is that a wide variety of clustering algorithms can be leveraged \cite{li2012twevent}. For example, community detection algorithms have been used in similar settings \cite{edouard2017graph, parikh2013events}. One of the most popular algorithms of this type is the Louvain method \cite{blondel2008fast}, which relies on modularity-based graph partitioning. Some key benefits of Louvain is that it is efficient on even large-scale networks and has a single parameter, resolution $R$, to tune.
\subsubsection{Cluster Linking}
\label{section:ClusterLinking}
Once we apply community detection to produce clusters for a given minute $C_{T}$, we link to the clusters from the previous minute $C_{T-1}$. We build a bipartite graph where the clusters in minute $T$ are on the right hand side and clusters in minute $T-1$ are on the left hand side. The edge weight between them is a measure of how many entities these clusters share, similar to the cosine similarity described earlier.

We filter out any edges whose weight falls below a threshold and perform maximum weighted bipartite matching \cite{kuhn1955hungarian} to find cluster links. When a cluster is successfully linked, we copy over the ID from the cluster in the previous minute onto the cluster in the current minute. For any clusters that are not linked, we generate a new, unique cluster ID. By linking clusters where appropriate, we form cluster chains.
\subsubsection{Cluster Ranking}
There are several ways to possibly rank clusters. Currently, we rank clusters based on the aggregate popularity of the entities contained within a cluster. As product use cases evolve, we would look to explore other ranking methods.
\subsubsection{Storage}
The linked, ranked list of clusters are persisted to internal stores such that they can be retrieved within the clustering service for future cluster linking steps or by other services.
\subsubsection{Parameter Tuning}
The key parameters are listed in Table ~\ref{table:keyparameters}. We observe that $S$ and $R$ have the most impact on clustering output, in terms of  coverage and quality, and are analyzed further in the Evaluation section. $W$ can be tuned as needed for memory reduction.

\begin{table}[h]
	\begin{tabular}{p{1.5cm}|p{4.5cm}}
	  Parameter & Description \\
		\hline
		Minimum similarity threshold $S$ & The minimum similarity threshold $S$ is applied to the edge weights of the entity graph. Drop edge weights below $S$. \\ \hline
		Louvain clustering resolution $R$ & The resolution is an important parameter for Louvain clustering. A larger resolution value will result in many smaller communities, and a smaller resolution value will result in few larger communities \cite{reichardt2006statistical}. \\ \hline
		Time window $W$ & Sliding window for aggregation of co-occurrences and frequencies.\\
		\hline
	\end{tabular}
	\caption{Summary of key system parameters}
	\label{table:keyparameters}
\vspace{-5mm}
\end{table}

\subsection{Algorithm}
We describe the psuedocode\footnotemark \ for the overall framework in Algorithm 1.
\begin{algorithm}[h]
	\caption{Similarity-Based Temporal Event Detection}
	\DontPrintSemicolon
	\SetAlgoLined
	\SetKwInOut{Input}{Input}\SetKwInOut{Output}{Output}
	\Input{$TweetStream$ a stream of tweets, $S$, minimum similarity threshold, $R$, resolution, $W$, time window}
	\Output{$L$, a list of clusters for a minute $T$}
	$M\leftarrow empty \ coOccurrence \ matrix$  \;
   	$Trends\leftarrow \{ set \ of \ trending \ entities$ \}  \;
   	\BlankLine
	\tcc{Running on background threads}
	\ForEach{Tweet in $TweetStream$}{
		$E\leftarrow extract \ each \ entity \ e \ from \ Tweet$ \\
		$Filtered\leftarrow filter(E,e \in Trends )$ \\
		\ForEach{Entity $E_f$ $\in$ Filtered}{
			updateCount($M$, $E_f$)\\
		}
	}
	\BlankLine
	\tcc{Each minute $T$, via a timer thread}
	remove($M$, $W$) \tcc{remove out-of-window updates}
	$G \leftarrow buildSimilarityGraph(M, S)$ \\
	$C_{T} \leftarrow Louvain(G, R)$\\
	$C_{T-1} \leftarrow fetch \ clusters \ for \ T-1$ \\
	$Links \leftarrow maxWeightedBipartiteMatching(C_{T}, C_{T-1})$ \\
	\ForEach{$c_t$ $\in$ $C_T$}{
		\uIf{($c_t$, $c_{t-1}$) $\in$ Links}{
			copy ID from $c_{t-1}$ to $c_t$ \\
		}
		$L \leftarrow L + c_t$
	}
	Sort $L$\\
	Return $L$
	\label{Algorithm1}
\end{algorithm}
\footnotetext{We are not able to share the code publicly, but the pseudocode is shown here for the purposes of reproducibility.}
\section{Evaluation}
\subsection{Evaluation Dataset}

\begin{table}[h]
\centering
  \begin{tabular}{l|c|p{2.5cm}|c}
  Entity & Chain ID & Title& Relevant? \\
  \hline
  \#madisonkeys & 1 & US Open Women's Quarterfinals& Y \\ \hline
  \#usopen & 1 & US Open Women's Quarterfinals & Y \\ \hline
  minnesota & 1 & US Open Women's Quarterfinals & N \\
  \hline
  \end{tabular}
  \caption{Examples of evaluation data}
  \label{table:EvaluationDatasetExample}
  \vspace{-0.8cm}
\end{table}

In the literature, we see that there is not a consistent benchmarking dataset by which each event detection system is measured. A large-scale evaluation corpus was created by McMinn et al. \cite{mcminn2013building}; however, it is most suitable for document-pivot methods.

To create an evaluation dataset, we start with one day’s English tweets from the United States. Three types of entities are extracted from these input tweets: hashtags, named entities, and internal knowledge graph based entities. Then we apply the end-to-end event detection process described above but without any similarity filtering. This allows us to produce a set of raw cluster chains and tune entity filtering processes in order to optimize cluster quality. For each cluster chain, we take all the entities from every point in time and produce one deduplicated set of entities per chain. For each chain, we select 20 representative tweets (10 most retweeted and 10 random tweets) that contained at least two co-occurring entities from the chain.

We manually examine representative tweets: if the chain corresponds to an event, we give it an ID and title (corresponding to the ``Chain ID'' and ``Title'' columns in Table \ref{table:EvaluationDatasetExample}). If the chain contains multiple events, we create different IDs and titles for each of them. Then we check all the titles and merge duplicates into single ID. We also mark and keep irrelevant entities as false positive examples. We present some examples of labeled data in Table \ref{table:EvaluationDatasetExample}. The dataset is  cross-validated by a separate individual to ensure reliability. In the end, our evaluation corpus contains 2695 entities and 460 events (i.e. different IDs). This labeling process requires manual effort but provides a valuable means to assess and improve system output.

\begin{figure*}[tb]
\centering
\subfloat[Events detected fraction]{\includegraphics[width=0.32\textwidth]{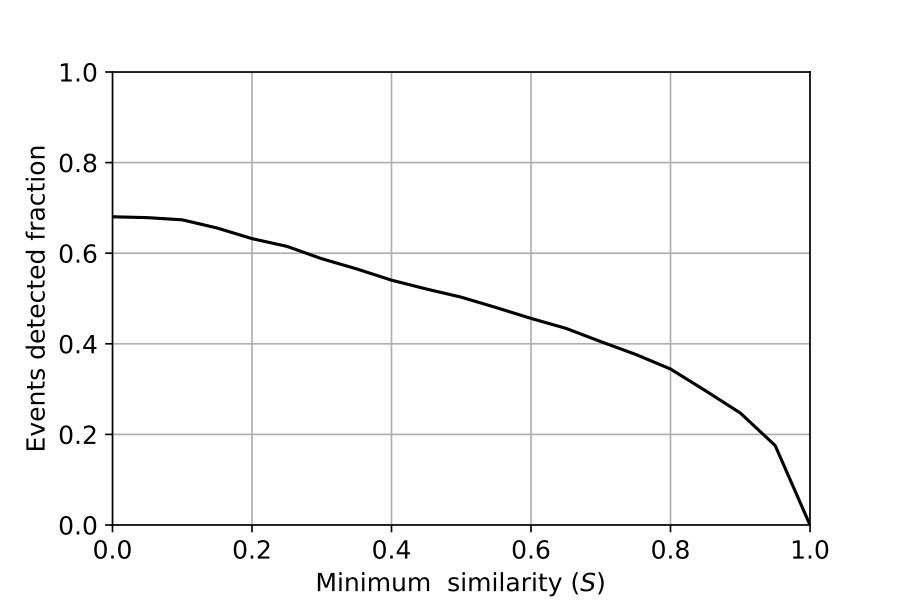}\label{fig:EventFraction}}
  \subfloat[Clustering quality (Connected Component)]{\includegraphics[width=0.32\textwidth]{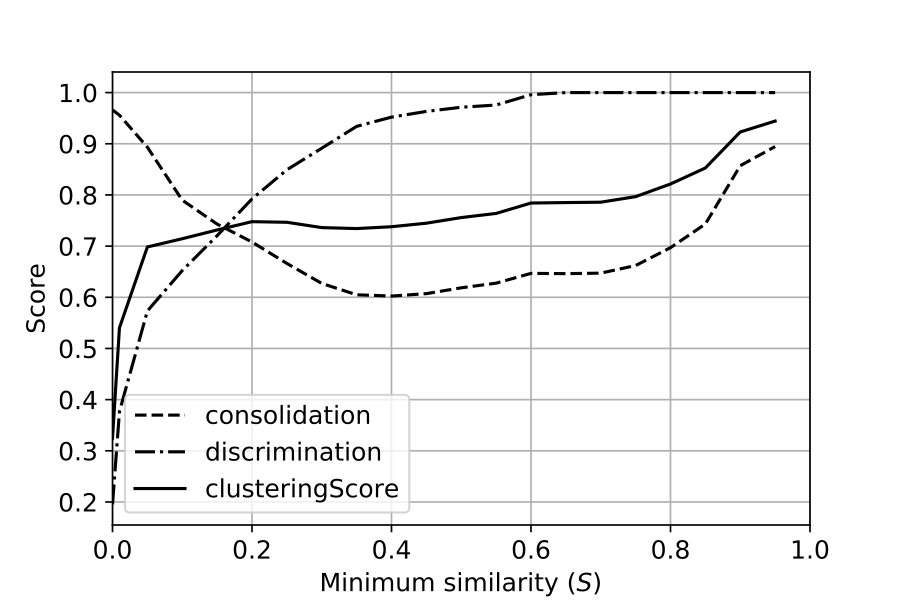}\label{fig:MinSimilarity}}
  \subfloat[Connected Component method vs. Louvain]{\includegraphics[width=0.32\textwidth]{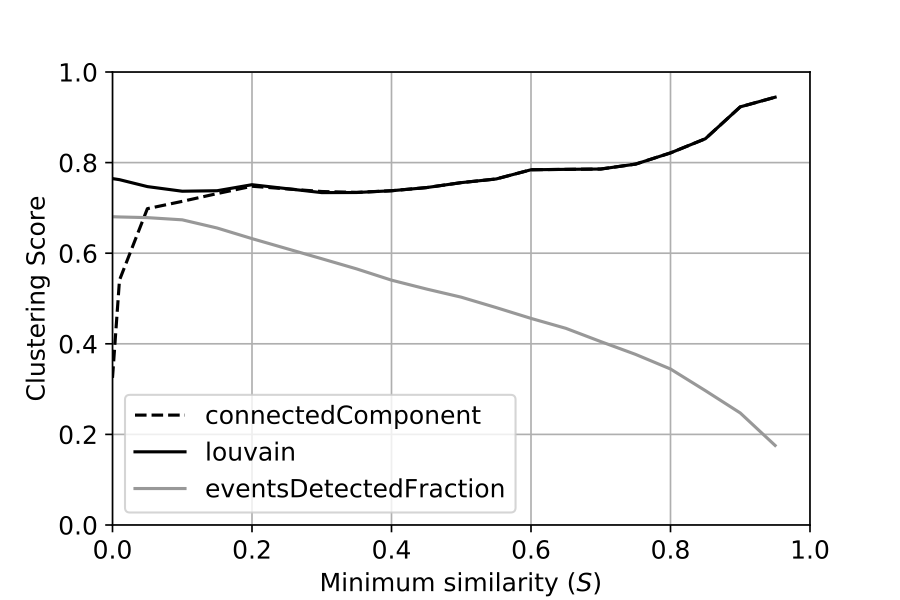}\label{fig:ConnectedComponentVsLouvain}}
\\
\subfloat[Louvain Resolution]{\includegraphics[width=0.32\textwidth]{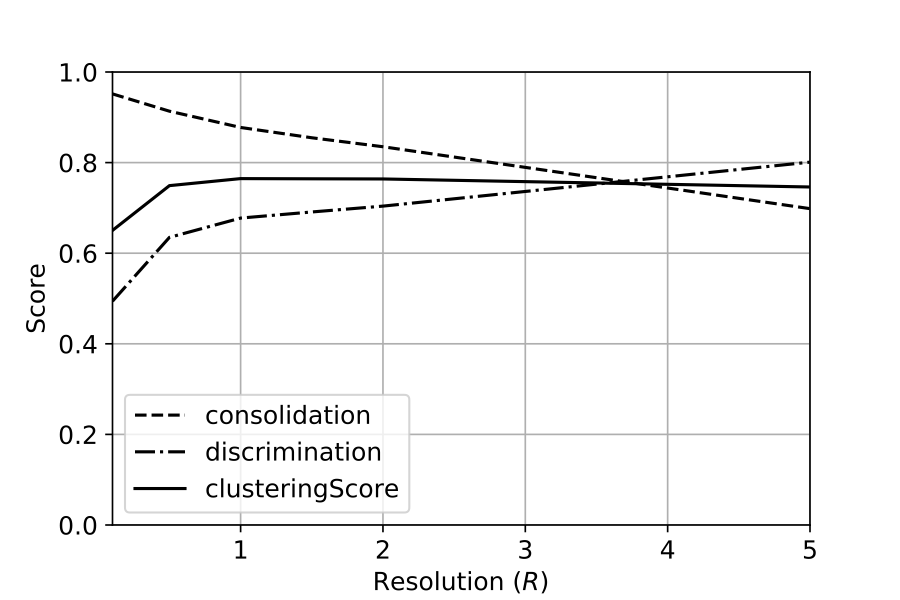}\label{fig:AppliedResolution}}
\subfloat[Merged event fraction]{\includegraphics[width=0.32\textwidth]{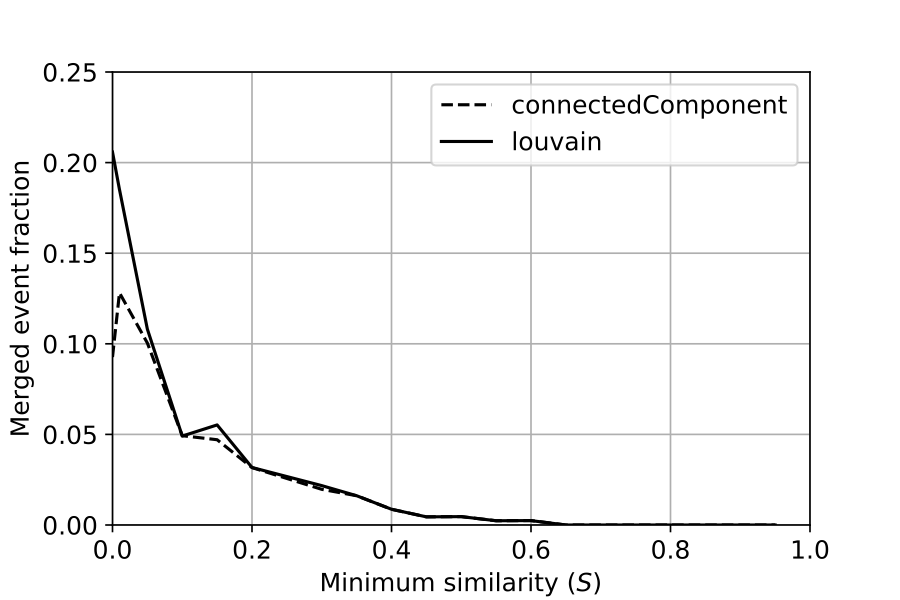}\label{fig:MergedEvent}}
\subfloat[Duplicate event fraction]{\includegraphics[width=0.32\textwidth]{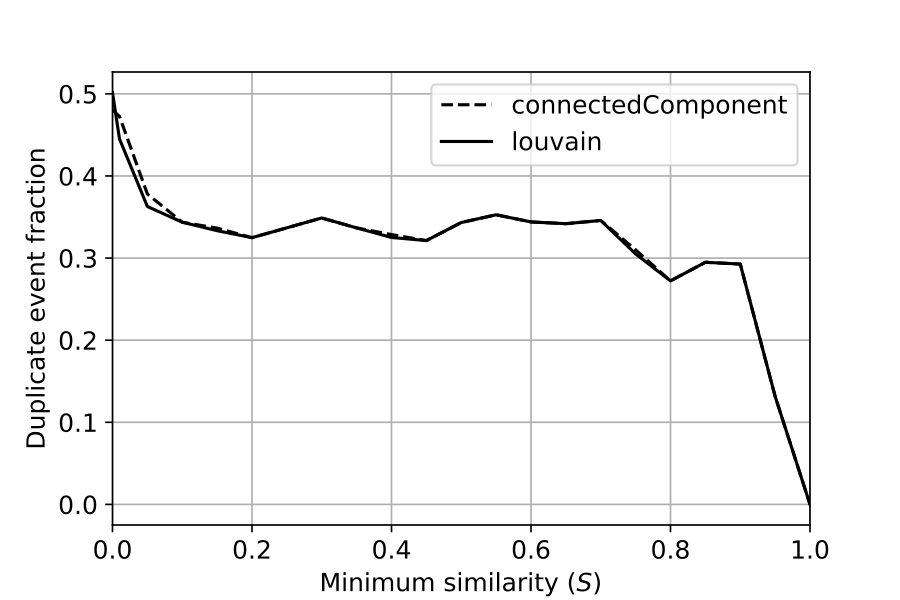}\label{fig:DuplicateEvent}}
  \caption{\textbf{System evaluation.} (a) Events detected fraction for different minimum similarities $S$; (b) The effects of minimum similarity $S$ on network structure; (c) Clustering score with Connected Component method vs. Louvain algorithm (with resolution $R=1$); (d) Clustering quality for different Louvain resolution $R$ settings; (e) Merged event fraction for different minimum similarities $S$; (f) Duplicate event fraction for different minimum similarities $S$.
  }
\end{figure*}

\subsection{Offline Evaluation}
In this section, we present the evaluation results for the proposed system. We run the system on the same set of tweets as the evaluation dataset but with different settings, and we measure their performance with the following metrics.

\subsubsection{Events detected fraction}
To start, we evaluate the coverage of our system. More specially, we compute the fraction of events from the evaluation set that are detected by our system. We define an entity as being unique if it is related only to one event. We consider an event to be detected if there exists a cluster of size greater than 1 that contains at least one unique entity of that event. Clustering quality is not the concern of this metric. Thus, if there exists a single cluster containing unique entities of several events, we consider all of those events to be detected. In our system, we aim to detect as many events as possible.

The minimum similarity filter for the graph is the primary filter that affects the fraction of events detected. In Figure \ref{fig:EventFraction}, we show events detected fraction for different minimum similarity $S$ settings. It is evident that increasing the minimum similarity setting decreases the fraction of events detected. This is due to more edges being filtered out; in other words, more nodes (entities) are isolated (i.e. without any edge) from the rest of the network and cannot be grouped into clusters. Note that even with a minimum similarity threshold of zero, our event detection fraction is less than 100\%. Given that we require that a cluster is comprised of at least two nodes, we do not include events from isolated nodes in the network.

\subsubsection{Clustering quality: Consolidation and Discrimination}
To assess quality, we have created a novel set of metrics. An important consideration when designing these metrics is that they do not penalize for detecting more events than the evaluation dataset nor for detecting more entities for an event. The new complementary metrics are called \textit{consolidation} and \textit{discrimination}, and they measure how effective we are at merging entities representing a single event and separating those of different events respectively. They are similar to the Rand index~\cite{rand1971objective} but allow us to assess mentioned aspects of clustering separately. Note that they also bear some resemblance to B-Cubed metrics~\cite{bagga1998entity}.

We mark two entities related if they are a part of single event in the ground truth and both of them are marked as relevant. We call two entities unrelated if they are a part of single event in the ground truth and exactly one of them is marked as irrelevant. We only consider those explicitly marked pairs because most of the entity pairs belonging to different events are very easy to distinguish. Therefore, we want to focus on what we call \emph{difficult} examples during our analysis.

Then, we define:
\begin{itemize}
  \item $t$: timestamp
  \item $T$: set of all timestamps in system output
  \item $A_t$: number of related entity pairs that are part of the system output at timestamp $t$
  \item $a_t$: number of related entity pairs that share a common cluster in the system output at timestamp $t$
  \item $B_t$: number of unrelated entity pairs that are part of the system output at timestamp $t$
  \item $b_t$: number of unrelated entity pairs that are not in a common cluster in the system output at timestamp $t$.
\end{itemize}

\textit{Consolidation} is defined as:
\begin{equation}
\textbf{C}=\frac{ \sum_{t \in t}{a_{t}}}{\sum_{t \in T}{A_{t}}}
\end{equation}

Similarly, \textit{discrimination} is defined as:
\begin{equation}
\textbf{D}=\frac{ \sum_{t \in t}{b_{t}}}{\sum_{t \in T}{B_{t}}}
\end{equation}

Intuitively, we can think of an algorithm putting all entities in a single cluster as achieving $100\%$ consolidation but $0\%$ discrimination. On the other hand, creating a cluster for each entity will yield $0\%$ consolidation and $100\%$ discrimination. It is important to optimize consolidation and discrimination together, much like it is important to optimize recall and precision in a machine learning system. We combine the two metrics into a single metric known as \textit{Clustering Score} using harmonic mean:
\begin{equation}
\textbf{CS}=(\frac{\textbf{C}^{-1}+\textbf{D}^{-1}}{2})^{-1}=\frac{2\textbf{C}\textbf{D}}{\textbf{C}+\textbf{D}}
\end{equation}

We can leverage these metrics to understand how the minimum similarity filter $S$ affects the network structure. In Figure \ref{fig:MinSimilarity}, we replace the clustering algorithm in our proposed system with a connected component detection method \cite{newman2003structure}. In this figure, we first notice that when the minimum similarity $S=0$, the consolidation $c=1$ and the discrimination $d=0$ since all nodes are connected (i.e. a complete graph). With the increase of minimum similarity, more edges are removed from the graph; thus the discrimination (dash-dot line) increases from 0 to 1.

The consolidation (dash line) is more interesting. When minimum similarity $S<0.4$, the increase of the minimum similarity filter value relates to lower consolidation, since more nodes are disconnected. But when $S>0.4$, most edges are removed, making many nodes isolated, resulting in them not being included in the final output. Remaining nodes are connected with heavy edges, and we achieve high consolidation as as a result. However, the size of clusters and the fraction of detected events tend to be very small.

In our system design, instead of relying on connected components, we use the Louvain community detection algorithm to achieve better clustering performance for minimum similarity $S<0.4$.  We compare the clustering score of the Louvain algorithm with the connected component method in Figure \ref{fig:ConnectedComponentVsLouvain}. It shows that when minimum similarity $S<0.2$, the Louvain algorithm achieves better performance because it successfully splits components into different clusters. When $S>0.2$, Louvain achieves the same results as the connected component method because the resulting components are too small to be split.

In figure \ref{fig:AppliedResolution}, we assess the performance of the Louvain algorithm with different resolution settings. Here we show the figure with minimum similarity $S=0.1$. It shows that $R=1$ is the setting that results in an optimal clustering score. Similar results were observed with other minimum similarity settings.

\subsubsection{Merged event fraction}
In addition to checking cluster quality and coverage, we also evaluate the cluster chains. Specifically, we check the fraction of chains that merge entities from different events. This metric is sensitive not only to clustering quality but also to the quality of cluster linking over time. Note that we only examine chains that last longer than 30 minutes for this evaluation. In Figure \ref{fig:MergedEvent}, we compare the merged event fraction for different minimum similarity settings. Based on the product use case, we select the appropriate threshold for minimum similarity, balancing the trade-off with respect to the total number of events detected.

\subsubsection{Duplicate event fraction}
The duplicate event fraction is an additional metric that we have designed. It is defined as the fraction of events in our evaluation dataset that have their entities identified in more than one chain. In Figure \ref{fig:DuplicateEvent}, we compare the duplicate event fraction for different minimum similarity settings. This number can be quite high. For example, minimum similarity $S=0.1$ results in about 35\% of events having duplicate chains. This may be high due to the fact that our evaluation dataset does not include sub-events. As we see in following case study section, some large events like the Golden Globes can have multiple sub-events, and it is more accurate to have them in different chains.

\subsection{Online Performance}
Below we profile the online system performance to demonstrate CPU and memory utilization under normal as well as atypical load scenarios. Most real-time event detection systems from the literature lack performance profiling; in some cases, they do so only in the context of a single large event \cite{gaglio2016framework}. We profile over a prolonged time range and demonstrate our system is able to scale and process millions of entities per minute.

\begin{figure}[h]
\vspace{-1mm}
\centering
\subfloat[Processed entities/sec]{\includegraphics[width=0.45\textwidth]{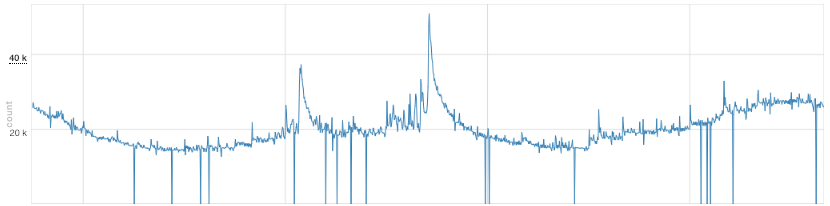}\label{fig:ProcessedEntities}}\\
\subfloat[CPU utilization profile]{\includegraphics[width=0.45\textwidth]{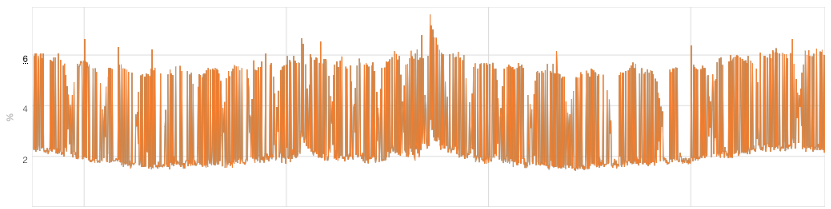}\label{fig:CPUProfile}}\\
\subfloat[Memory usage profile]{\includegraphics[width=0.45\textwidth]{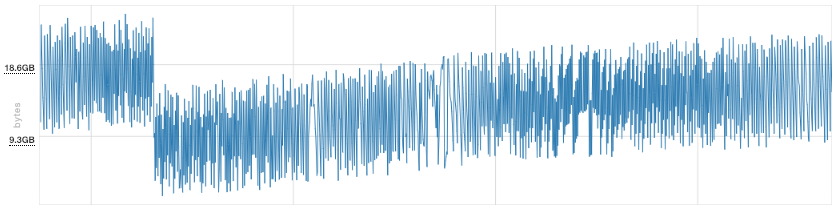}\label{fig:GCProfile}}\\
\caption{Performance over a time range. Note that time scale is omitted due to proprietary data restrictions.}
\end{figure}

As seen in Figure \ref{fig:CPUProfile}, the CPU usage is typically low (< 10\%) and consistent over a normal day’s traffic. Similarly in Figure \ref{fig:GCProfile}, memory usage is consistent. The system is deployed on a Java Virtual Machine (JVM) and thus relies on Garbage Collection (GC) for memory management. The decrease in memory usage near the start of the graph depicts a major collection, which occur only once every 1-2 days due to the stability of long-lived objects. Minor collections steadily occur every 1-2 minutes.

We observe one particular instance at the center of Figure \ref{fig:ProcessedEntities} where we handle spikes of up to $50K$ processed entities per second (PSEC). During this load spike, which represents a doubling of PSEC in a short period of time, the corresponding CPU increase and memory impact as seen in the figures below is negligible.

Thus, the system exhibits stable CPU and memory usage even when faced with abnormal load. We achieve this in our implementation through various means such as employing load shedding techniques, leveraging memory efficient data structures, minimizing object churn, and taking advantage of sparsity in computations when possible.

When we encounter spikes in entities to process, it is usually due to one of the following scenarios:
\begin{enumerate}
	\item \emph{Subscriber lag} - The stream we are subscribing to slows down unexpectedly, leading to a backlog of unprocessed tweets
	\item \emph{Bursty traffic} - Usage of the platform can spike for indeterminate periods of time, leading to a sharp rise in tweets ingested
	\item \emph{Downstream lag} - The rate of tweet processing within the system can slow due to lag in a downstream system required for tweet processing, leading to a backlog of unprocessed tweets.
\end{enumerate}

These symptoms can result in atypical load, and we have to adjust with the appropriate amount of temporary load shedding. The goal is to gracefully degrade while ensuring service uptime. Below is an example of doing so in the face of subscriber lag:

\begin{figure}[h]
	\includegraphics[width=0.4\textwidth]{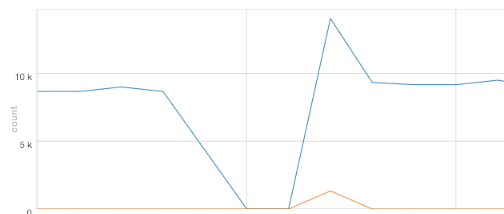}
	\caption{Load shedding}
	\label{fig:LoadShedding}
\end{figure}

In Figure \ref{fig:LoadShedding}, the top line represents tweets processed per second and the bottom line represents dropped tweets per second. Note that the shedding is temporary (only takes place during a given minute) and the system resumes to normalcy afterwards. Safeguards like this were important in order to improve the resiliency of the system.

\subsection{Other evaluation methods}
\begin{figure}[h]
	\includegraphics[width=0.2\textwidth,keepaspectratio]{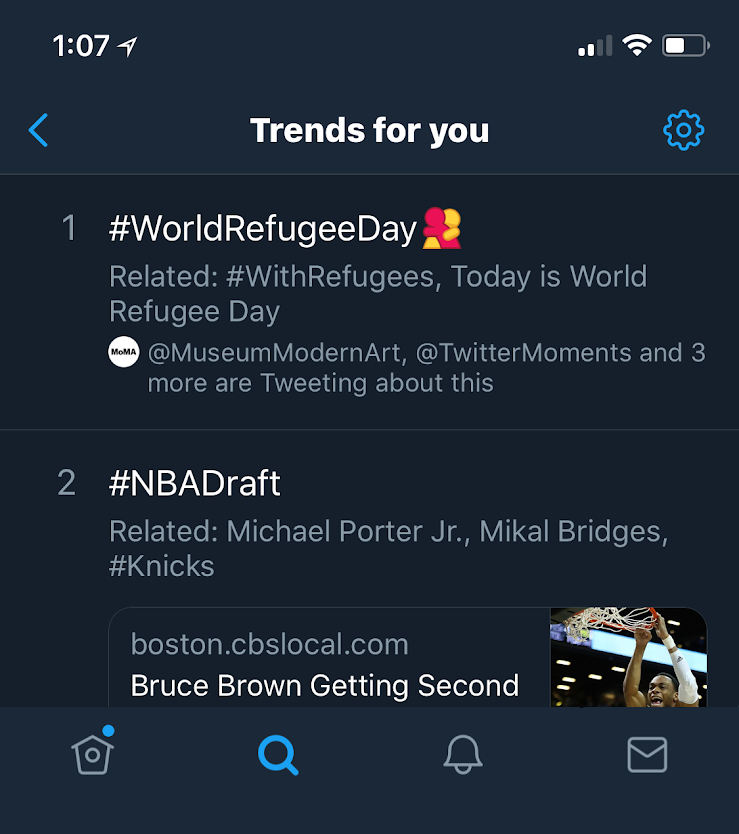}
	\caption{Trend folding on Twitter. Related trending entities are shown via the ``Related'' line.}
	\label{fig:TrendFolding}
\end{figure}

In addition to the procedures described above, we have also performed other types of evaluations:
\begin{enumerate}
	\item \emph{Live system output monitoring} - We have been manually reviewing the system output, especially during important events, since launch. This has allowed us to spot edge cases not observed during offline evaluation and also has given us a sense as to how the numbers from offline evaluation translate to clustering quality in real world scenarios and what trade-offs are most desirable.
	\item \emph{Periodic human computation evaluation} - Every week, the top clusters, based on product usage, are sent to human annotators for manual quality checking. A cluster is judged as ``good'' if at least 90\% of the entities constituting that cluster are judged as relevant to that cluster. Over 95\% of the clusters shown in the United States in a recent assessment were judged as good. In addition to validating quality, human annotation could potentially transform our research into a supervised learning problem and provide data to continuously improve the system quality.
	\item \emph{A/B testing} - The described event detection system has been used for Trends folding on Twitter (Figure \ref{fig:TrendFolding}). We group related trending entities from the same detected event in order to give users more context. Our A/B testing showed that with this feature enabled, users gained a better understanding of the Trends shown and were more likely to interact with them. This demonstrates inherent user value in being able to detect and contextualize events.
\end{enumerate}

\section{Case Study: 76TH GOLDEN GLOBE AWARDS}
The Golden Globe Awards are one of the most important awards in the film industry. The 76th annual ceremony was held on January 6, 2019 starting at 01:00 UTC. It is an example of an important event with a great degree of conversation on Twitter. In this section, we present how our system performed during that event and how accurately it reflected real world conversation. The temporal aspect of the event is of particular interest as we examine how entity clusters emerge, evolve, and disappear.

\begin{figure}[h]
	\includegraphics[width=\columnwidth,keepaspectratio]{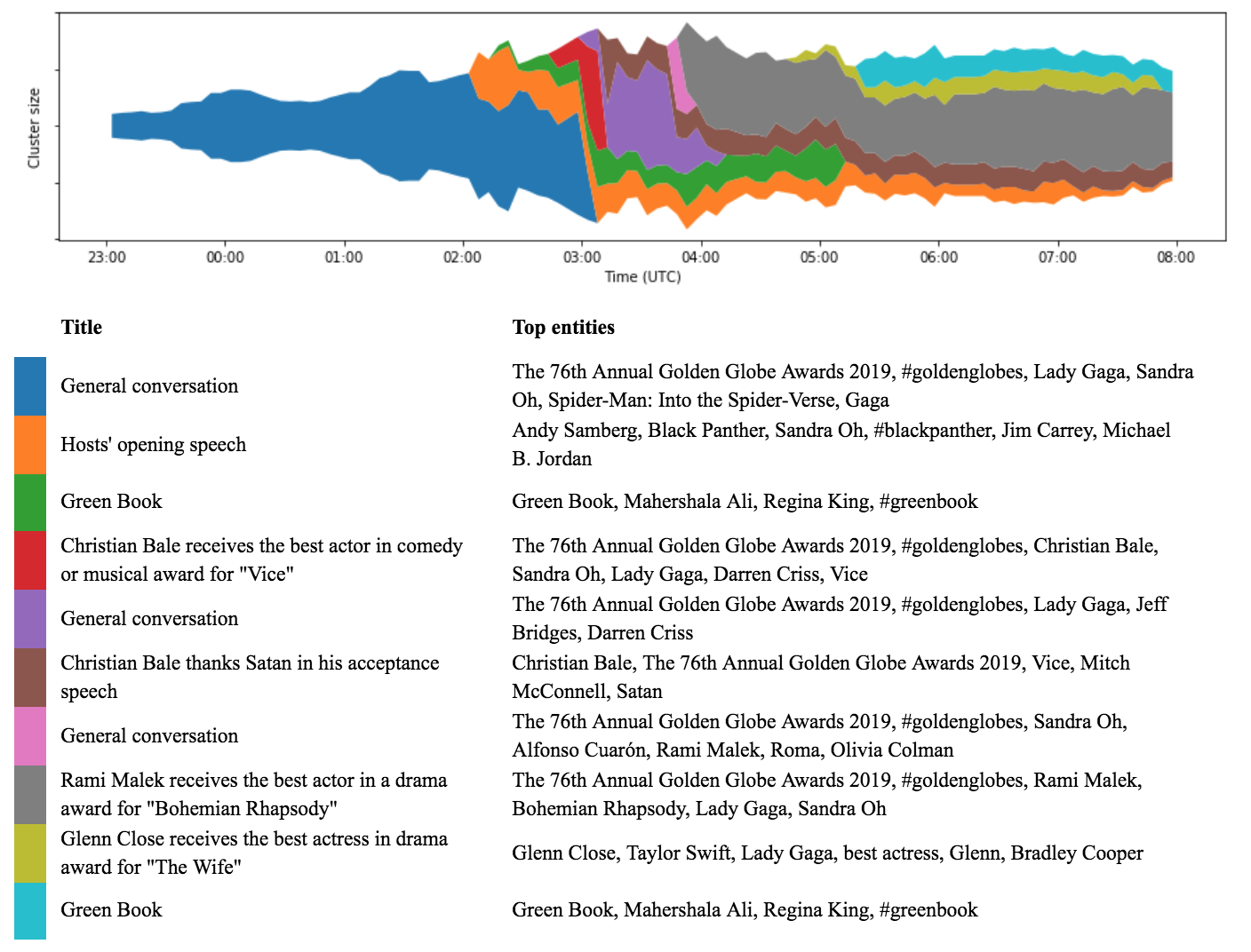}
	\caption{Top cluster chains related to Golden Globes over time. Stream height corresponds to number of entities in a cluster.}
	\label{fig:Oscars}
\end{figure}

Figure \ref{fig:Oscars} is an overview of the event: it shows the ten biggest cluster chains (in terms of total tweet count) that at any time between January 6th and 7th contain an entity matching ``*golden*globes*'' pattern where ``*'' represents any, possibly empty, sequence of characters. In the figure, we can see that the event structure is stable outside the main ceremony time, i.e., before 01:00 UTC and after 04:30 UTC. Before the ceremony, all related entities are clustered into a single big chain since no particular theme has yet emerged. After the ceremony, users continue to talk about topics that were popular during the ceremony (e.g. Green Book, Glenn Close, and Rami Malek). However, the most interesting period is during the ceremony itself: the system is able to capture fast evolving topics and create different chains for them.

\begin{figure}[h]
	\includegraphics[width=\columnwidth,keepaspectratio]{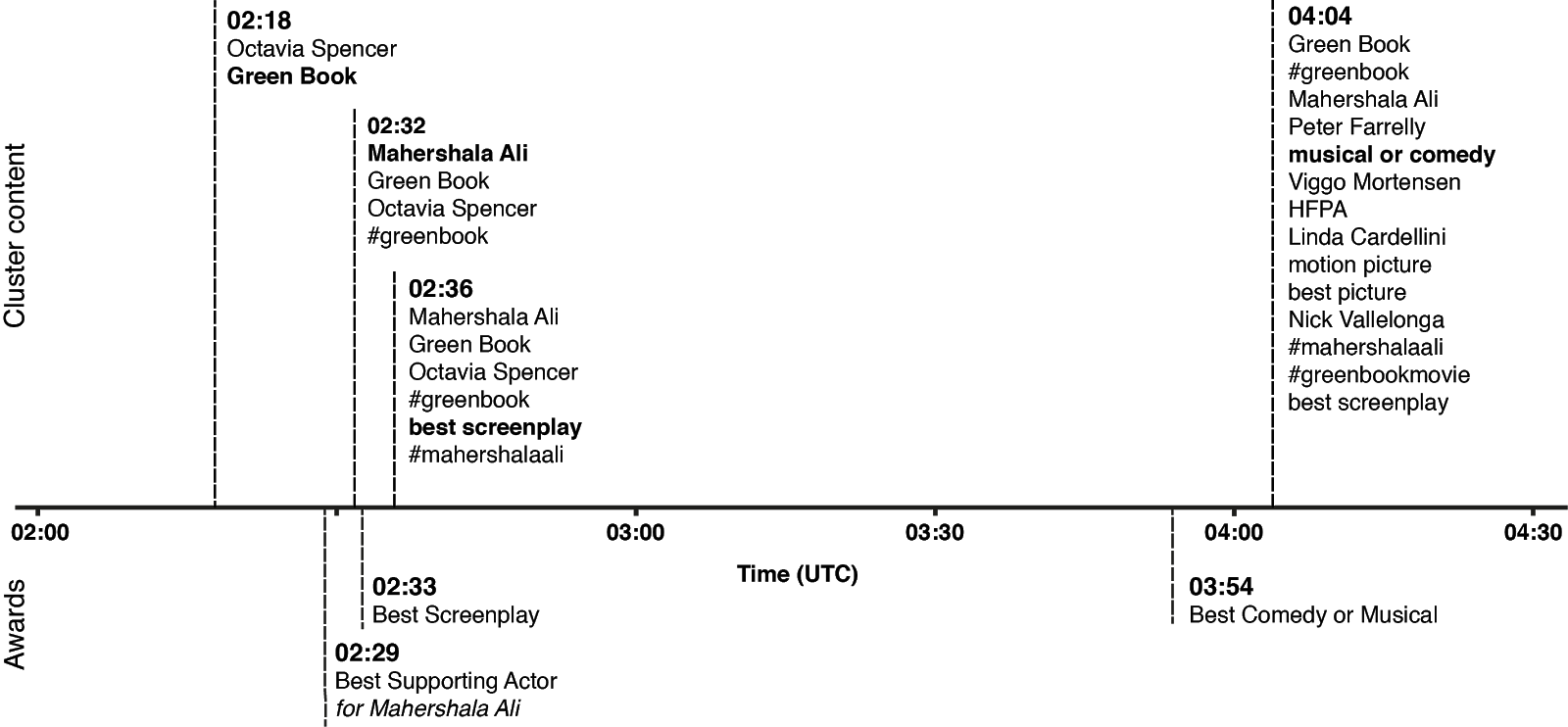}
	\caption{Green Book cluster evolution and corresponding award presentations times.
	}
	\label{fig:GreenBook}
\end{figure}

\emph{Green Book} was the most acclaimed movie: it was awarded Best Supporting Actor, Best Screenplay, and Best Musical or Comedy. In Figure \ref{fig:GreenBook}, we show in detail how the Green Book chain (also represented as the green area of Figure \ref{fig:Oscars}) evolves over time. At first, it only consists of entities representing the movie and its executive producer (Olivia Spencer). Shortly after, Mahershala Ali received his award for Best Actor, and the appropriate entity is added to the cluster.
Similarly, the ``best screenplay'' entity is added thereafter. When the Best Musical or Comedy Award is presented, the cluster represents the fully developed conversation: related hashtags are used and film crew members are mentioned. At 05:09 UTC, a new chain emerged out of the original Green Book chain, represented as light blue and green areas in Figure \ref{fig:Oscars} respectively. Subsequently at 05:12 UTC, the new chain absorbed the old one while retaining the new ID. Proper representation of such behavior is under investigation.

The retrospective analysis that we have performed is interesting from a sociological perspective in that it provides insights about what was important to people during the event. Moreover, as these charts can be generated in real time, they can be used for tracking real world events as they unfold.

\section{Conclusion}
In this paper, we have presented an event detection method that is able to handle event evolution over time and was deployed to work in real time at Twitter scale. We described how it is designed and how we evaluated its performance, both offline and online. To measure and maintain quality over time, we perform continuous evaluation via human annotators. Product application is the primary driver during all development phases. One of the first use cases was folding trends in official Twitter clients (Figure \ref{fig:TrendFolding}). A/B testing validated its positive impact on user experience. Internally, the system is also used for event discovery by Twitter curators who track the most important trends on the platform and provide curated collections of tweets. 

Other possible applications include timeline ranking and search query expansion. Further research and system quality improvements will be motivated by subsequent use cases. For example, instead of relying on entity co-occurrence, we could compare their contexts modeled as embeddings. As events represent something atypical, precomputed embeddings may not be suitable. Instead, they need to be dynamically updated; this presents an interesting challenge for future investigation.

\begin{acks}
The authors thank Guy Hugot-Derville for his guidance and his initial drive to research this idea. We also thank Alexandre Lung-Yut-Fong, Jeff de Jong, Samir Chainani, Lei Wang, David Blackman, Gilad Buchman, Prakash Rajagopal, Volodymyr Zhabiuk, Lu Gao, Dennis Zhao, Ashish Bansal, and Ajeet Grewal for their support.
\end{acks}

\bibliographystyle{ACM-Reference-Format}
\balance
\bibliography{bib/kdd_evd}
\end{document}